\documentclass[preprint,12pt]{elsarticle}
\makeatletter
\def\ps@pprintTitle{%
 \let\@oddhead\@empty
 \let\@evenhead\@empty
 \def\@oddfoot{\centerline{\thepage}}%
 \let\@evenfoot\@oddfoot}
\makeatother

\usepackage{graphicx}
\usepackage{float}
\usepackage{amsmath}
\usepackage{bm}
\usepackage{amstext}
\usepackage{optidef}
\usepackage{xcolor}
\usepackage{algorithm}
\usepackage{algorithmic}
\usepackage{amssymb}
\usepackage[colorlinks]{hyperref}
\usepackage{dirtree}
\usepackage{forest}
\usepackage{array}
\usepackage{tabu}
\usepackage{url}
\usepackage{listings}
\usepackage{enumitem}
\usepackage{times}
\usepackage{tgtermes}
\usepackage{lmodern}
\usepackage[T1]{fontenc}
\usepackage{multirow}
\usepackage{longtable}
\usepackage{subfigure}
\usepackage[left=2cm,right=2cm,asymmetric]{geometry}

\definecolor{folderbg}{RGB}{124,166,198}
\definecolor{folderborder}{RGB}{110,144,169}
\definecolor{mygreen}{RGB}{28,172,0} 
\definecolor{mylilas}{RGB}{170,55,241}
\definecolor{codegreen}{rgb}{0,0.6,0}
\definecolor{codegray}{rgb}{0.5,0.5,0.5}
\definecolor{codepurple}{rgb}{0.58,0,0.82}
\definecolor{backcolour}{rgb}{0.95,0.95,0.92}

\lstdefinestyle{mystyle}{
    backgroundcolor=\color{backcolour},   
    commentstyle=\color{codegreen},
    keywordstyle=\color{magenta},
    numberstyle=\tiny\color{codegray},
    stringstyle=\color{codepurple},
    basicstyle=\footnotesize,
    breakatwhitespace=false,         
    breaklines=true,                 
    captionpos=b,                    
    keepspaces=true,                 
    numbers=left,                    
    numbersep=5pt,                  
    showspaces=false,                
    showstringspaces=false,
    showtabs=false,                  
    tabsize=2
}
 
\def\Size{4pt}
\tikzset{
      folder/.pic={
        \filldraw[draw=folderborder,top color=folderbg!50,bottom color=folderbg]
          (-1.05*\Size,0.2\Size+5pt) rectangle ++(.75*\Size,-0.2\Size-5pt);  
        \filldraw[draw=folderborder,top color=folderbg!50,bottom color=folderbg]
          (-1.15*\Size,-\Size) rectangle (1.15*\Size,\Size);
      }
    }

\journal{Journal Name}

\begin{document}

\sloppy

\begin{frontmatter}


\title{PhasePack User Guide \\ {\small \url{https://github.com/tomgoldstein/phasepack-matlab}}}
\author{Rohan Chandra, Ziyuan Zhong, Justin Hontz, Val McCulloch \\
Christoph Studer, Tom Goldstein }



\end{frontmatter}


\tableofcontents

\section{Introduction \& Motivation }
\label{S:1}

Phase retrieval \cite{fienup1978reconstruction,jaganathan2015phase} finds important applications in many physical problems in optical imaging \cite{millane1990phase,bian2015fourier}, quantum state tomography, electron microscropy \cite{misell1973method}, X-ray crystallography, astronomy \cite{fienup1982phase} and X-ray diffraction imaging \cite{shechtman2015phase}. The goal is to retrieve a complex-valued signal $\bm{x}\in \mathbb{C}^n$ from measurements of the form 
\begin{equation}
\bm b = |A \bm x|, \label{pr}
\end{equation}
where $|A \bm x|$ denotes the magnitudes of the (possibly complex-valued) linear measurements $A\bm x.$ 

While there has been a recent explosion in development of phase retrieval methods, the lack of a common interface has made it difficult to compare new methods against the current state-of-the-art. 
The purpose of PhasePack is to create a common interface for a wide range of phase retrieval schemes, and to provide a common test bed using both synthetic data and empirical imaging datasets \cite{metzler2017coherent}. 


%
%
%
%
%
%
%

\section{What is Phasepack?}
Phasepack provides Matlab implementations for both non-convex and convex algorithms that solve the phase retrieval problem. An algorithm list can be found in section (\ref{S:22}). Phasepack benchmarks many recent methods against one another and generates performance comparisons with varying numbers of samples, SNR (signal-to-noise ratio), iterations, and runtime. The package handles single method testing as well as multiple method comparisons.\\ 

\begin{figure}[H]
\centering
\includegraphics[width=\linewidth/2, trim=0cm 8cm 0cm 8.0cm, clip]{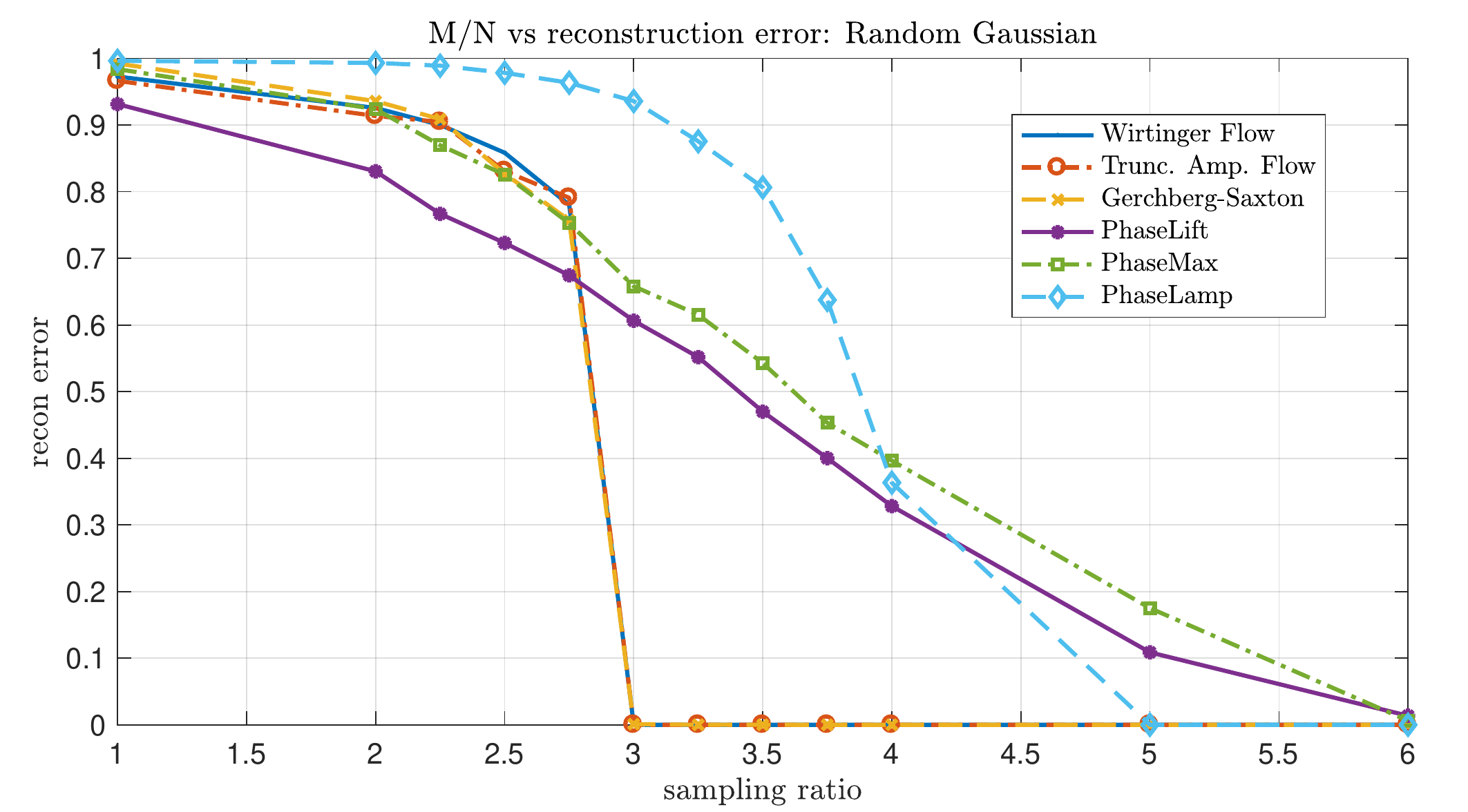}~
\includegraphics[width=\linewidth/2, trim=0cm 8cm 0cm 8.0cm, clip]{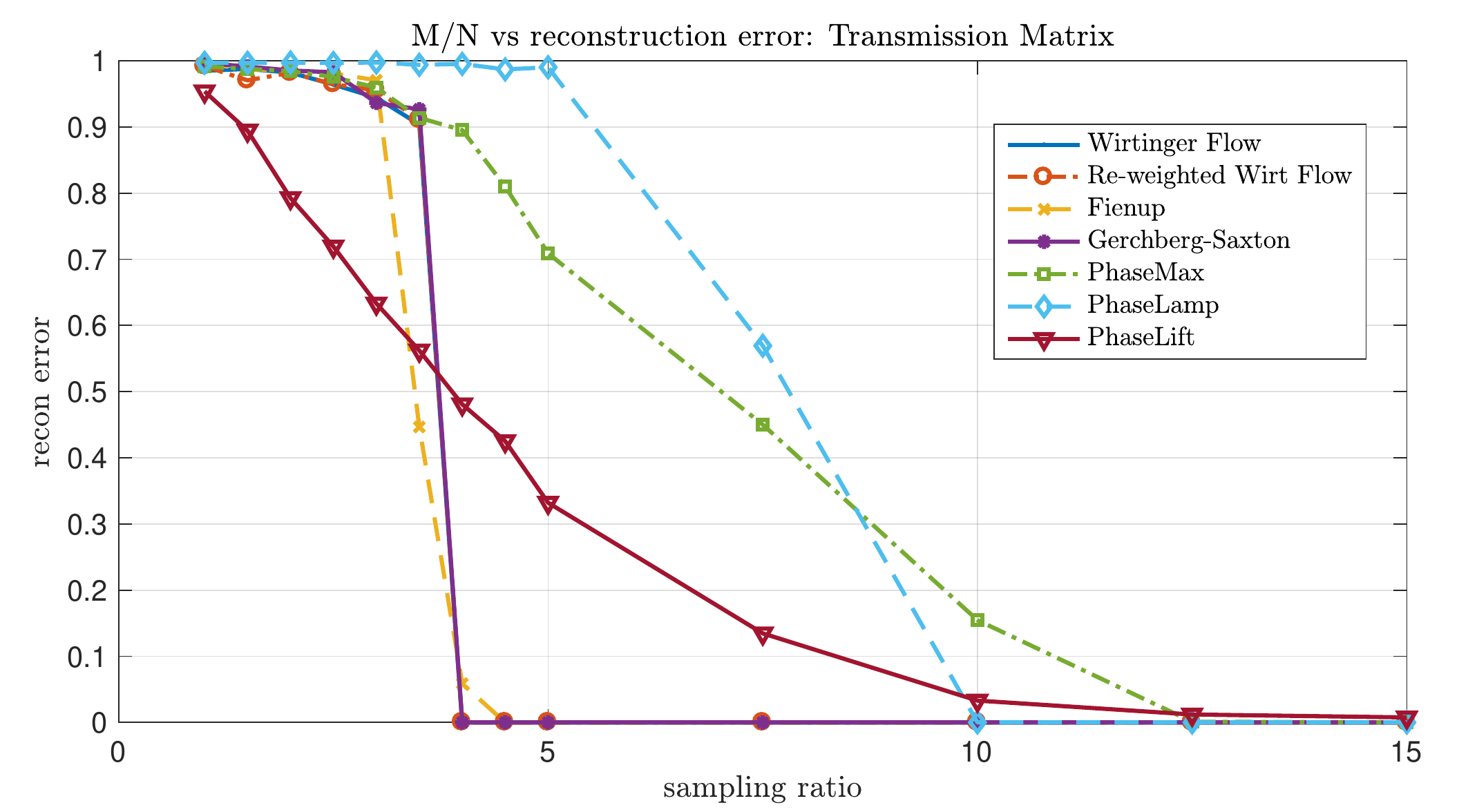}
\caption{ Reconstruction error as a function of sampling ratio (number of measurements over number of unknowns) for a few algorithms. (left)
Signal of length 100 using synthetic Gaussian measurements. (right) Signal of length 256 using an empirical transmission matrix measurement operator.}
\label{fig:random}
\end{figure}

\begin{figure}[H]
\centering
\begin{minipage}{3cm} \centering
\vspace{12pt}Original\vspace{14pt}
\includegraphics[width=3cm]{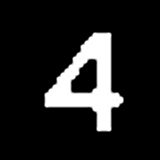}

\end{minipage}
\begin{minipage}{3cm} \centering
\vspace{12pt}Fienup \\
0.21125\vspace{3pt}
\includegraphics[width=3cm]{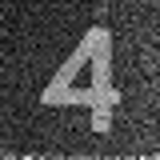}
\end{minipage}
\begin{minipage}{3cm} \centering
Gerchberg-Saxton\\0.21125\vspace{1pt}
\includegraphics[width=3cm]{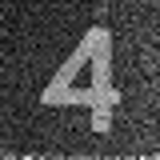}
\end{minipage}
\begin{minipage}{3cm} \centering
\vspace{12pt}Wirtinger Flow\\0.22900\vspace{3pt}
\includegraphics[width=3cm]{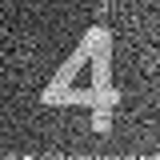}
\end{minipage}
\begin{minipage}{3cm} \centering
Truncated Wirtinger Flow\\0.53715\vspace{1pt}
\includegraphics[width=3cm]{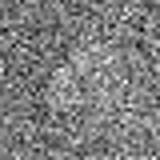}
\end{minipage}

\begin{minipage}{3cm} \centering
\vspace{3pt}Reweighted Wirtinger Flow\\0.22045\vspace{3pt}
\includegraphics[width=3cm]{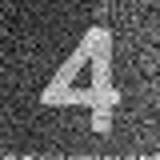}
\end{minipage}
\begin{minipage}{3cm} \centering
\vspace{17pt}Amplitude Flow\\0.21125\vspace{3pt}
\includegraphics[width=3cm]{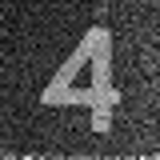}
\end{minipage}
\begin{minipage}{3cm} \centering
\vspace{3pt}Truncated Amplitude Flow\\0.49096\vspace{3pt}
\includegraphics[width=3cm]{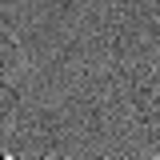}
\end{minipage}
\begin{minipage}{3cm} \centering
\vspace{3pt}Reweighted Amplitude Flow\\0.21800\vspace{3pt}
\includegraphics[width=3cm]{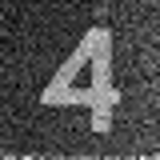}
\end{minipage}
\begin{minipage}{3cm} \centering
\vspace{17pt}Kaczmarz\\0.78187\vspace{3pt}
\includegraphics[width=3cm]{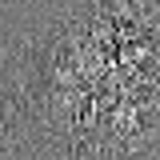}
\end{minipage}
\begin{minipage}{3cm} \centering
\vspace{3pt}Coordinate Descent\\0.43956\vspace{3pt}
\includegraphics[width=3cm]{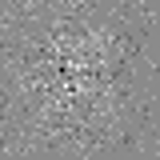}
\end{minipage}
\begin{minipage}{3cm} \centering
\vspace{17pt}PhaseLift\\0.98754\vspace{3pt}
\includegraphics[width=3cm]{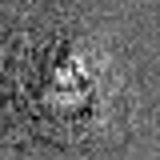}
\end{minipage}
\begin{minipage}{3cm} \centering
\vspace{17pt}PhaseMax\\0.72140\vspace{3pt}
\includegraphics[width=3cm]{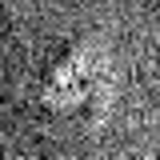}
\end{minipage}
\begin{minipage}{3cm} \centering
\vspace{17pt}PhaseLamp\\0.75492\vspace{3pt}
\includegraphics[width=3cm]{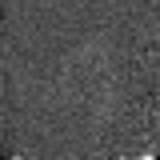}
\end{minipage}
\begin{minipage}{3cm} \centering
\vspace{17pt}SketchyCGM\\0.92215\vspace{3pt}
\includegraphics[width=3cm]{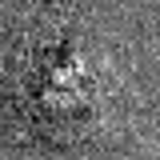}
\end{minipage}
\caption{ Reconstruction of a 40x40 image using empirical phaseless measurements. Below each algorithm name is 
the relative measurement error ($\||Ax|-b\|/\|b\|$) achieved by each method.} 
\label{fig:emp}
\end{figure}

\begin{figure}[H]
\centering
\begin{minipage}{3cm} \centering
\vspace{12pt}Original\vspace{14pt}
\includegraphics[width=3cm]{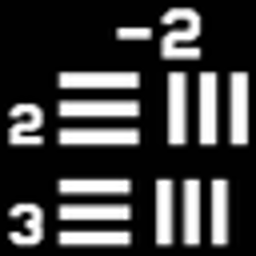}

\end{minipage}
\begin{minipage}{3cm} \centering
\vspace{12pt}Fienup \\
0.13019\vspace{3pt}
\includegraphics[width=3cm]{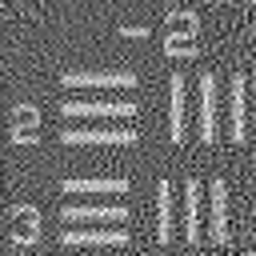}
\end{minipage}
\begin{minipage}{3cm} \centering
Gerchberg-Saxton\\0.13019\vspace{3pt}
\includegraphics[width=3cm]{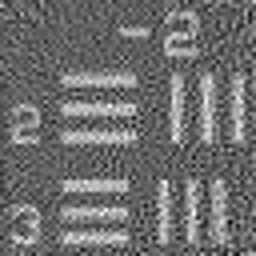}
\end{minipage}
\begin{minipage}{3cm} \centering
\vspace{12pt}Wirtinger Flow\\0.22900\vspace{3pt}
\includegraphics[width=3cm]{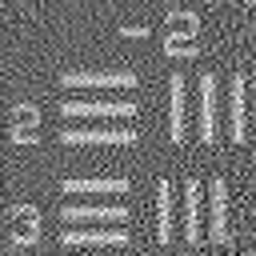}
\end{minipage}
\begin{minipage}{3cm} \centering
Truncated Wirtinger Flow\\0.18017\vspace{3pt}
\includegraphics[width=3cm]{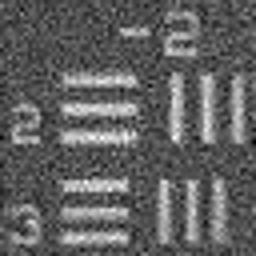}
\end{minipage}

\begin{minipage}{3cm} \centering
\vspace{3pt}Reweighted Wirtinger Flow\\0.14044\vspace{3pt}
\includegraphics[width=3cm]{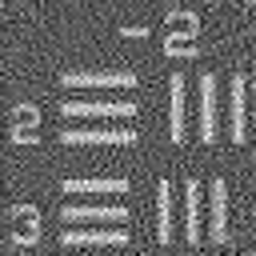}
\end{minipage}
\begin{minipage}{3cm} \centering
\vspace{17pt}Amplitude Flow\\0.13017\vspace{3pt}
\includegraphics[width=3cm]{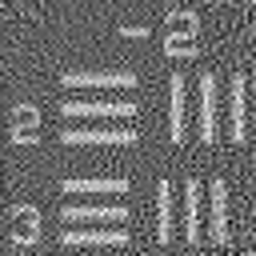}
\end{minipage}
\begin{minipage}{3cm} \centering
\vspace{3pt}Truncated Amplitude Flow\\0.43330\vspace{3pt}
\includegraphics[width=3cm]{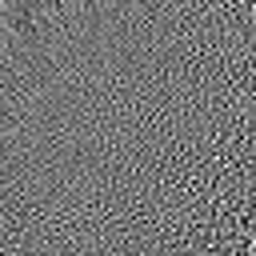}
\end{minipage}
\begin{minipage}{3cm} \centering
\vspace{3pt}Reweighted Amplitude Flow\\ 0.13261\vspace{3pt}
\includegraphics[width=3cm]{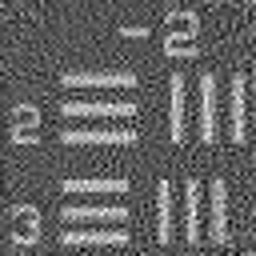}
\end{minipage}
\begin{minipage}{3cm} \centering
\vspace{17pt}Kaczmarz\\0.34500\vspace{3pt}
\includegraphics[width=3cm]{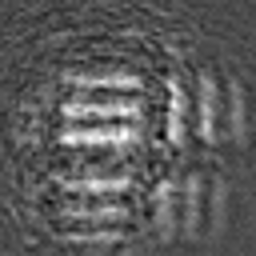}
\end{minipage}
\begin{minipage}{3cm} \centering
\vspace{3pt}Coordinate Descent\\0.21616\vspace{3pt}
\includegraphics[width=3cm]{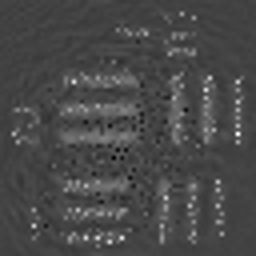}
\end{minipage}
\begin{minipage}{3cm} \centering
\vspace{17pt}PhaseLift\\0.35452\vspace{3pt}
\includegraphics[width=3cm]{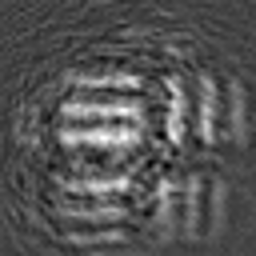}
\end{minipage}
\begin{minipage}{3cm} \centering
\vspace{17pt}PhaseMax\\0.23459\vspace{3pt}
\includegraphics[width=3cm]{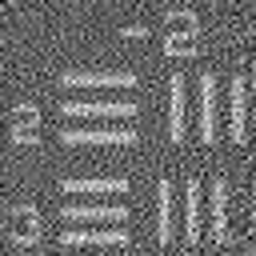}
\end{minipage}
\begin{minipage}{3cm} \centering
\vspace{17pt}PhaseLamp\\0.64980\vspace{3pt}
\includegraphics[width=3cm]{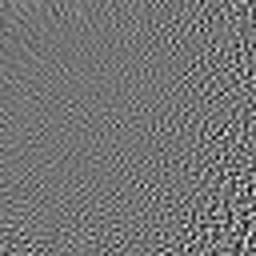}
\end{minipage}
\begin{minipage}{3cm} \centering
\vspace{17pt}SketchyCGM\\0.85526\vspace{3pt}
\includegraphics[width=3cm]{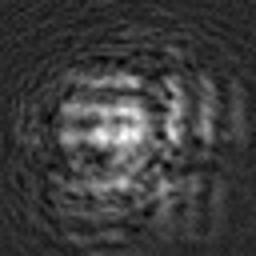}
\end{minipage}
\caption{ Reconstruction of a 64x64 image (modified air force target) using empirical phaseless measurements. Below each algorithm name is 
the relative measurement error ($\||Ax|-b\|/\|b\|$) achieved by each method.} 
\label{fig:aft}
\end{figure}

Examples of experiments generated by PhasePack are shown below.  All of these examples were produced using the scripts included in the \path{benchmark} sub-folder of the PhasePack distribution.  Figure \ref{fig:random} shows the performance of several algorithms for recovering a random signal from random Gaussian measurements (left) and from measurements acquired using a real empirical transmission matrix (right).  Figures \ref{fig:emp} and \ref{fig:aft} show the reconstruction of an image from empirical measurements obtained using an optical device \cite{metzler2017coherent}.  See Section \ref{sec:tm} for details on how to obtain and use the empirical datasets.




\section{PhasePack at a glance}
\label{S:13}

The \path{PhasePack} folder contains several sub-folders, and two top-level files.  The file \path{runSignalReconstructionDemo} is a simple script that demonstrates how to build a phase retrieval problem and solve it using PhasePack.   The script \path{runImageReconstructionDemo} does the same, but with a more complex Fourier measurement operator, and with a 2d image instead of a 1d signal.  {\em Note: any file whose name begins with ``run...'' is a demo script that can be run ``out of the box'' with no arguments.}

Descriptions of the contents of each sub-folder within PhasePack are given below.

\begin{itemize}[leftmargin=9em]
\item[\path{examples}]  Demo scripts that show how to setup and run each algorithm.  Separate scripts exists for each phase retrieval algorithm, and for each initialization method.   For example, the script \path{runGerchbergSaxton} sets up a phase retrieval problem, and solves it using the classical Gerchberg-Saxton routine.  The script \path{runInitOrthogonal} demonstrates how to use the ``orthogonality-promoting'' (aka ``null'') initializer to generate an initial guess.
\item[\path{benchmarks}]  Scripts for comparing multiple algorithms and plotting performance curves for different test problems.  For example, the script \path{runBenchmark1DGaussianMN} plots the accuracy of reconstruction methods as a function of the data sampling ratio ($m/n,$ the number of measurements over the signal length) when using a random Gaussian measurement matrix.  Other routines exist for using various empirical and synthetic test problems.
\item[\path{solvers}]  Implementations of individual phase retrieval methods, and supporting files that perform general gradient methods (e.g., L-BFGS) and line search.
\item[\path{initializers}]  Initialization algorithms.  This folder contains a range of spectral methods for producing approximate solutions to phase retrieval problems.
\item[\path{benchmarkResults}]  Results produced by scripts in the \path{benchmark} folder are stored here. This mostly consists of images that are saved after reconstruction.
\item[\path{util}]  Assorted methods for plotting results, handling data structures, managing user-supplied options and their default value, etc... 
\item[\path{data}]  Data files used for phase retrieval experiments.  Some benchmarks require the user to download large datafiles that do not come with PhasePack, and place these data files in the \path{data} folder.  See Section \ref{sec:tm}.
\end{itemize}

\section{Using a Solver}
\label{S:2}

A particular phase retrieval algorithm and initializer can be used by invoking the general interface as follows.

\begin{lstlisting}[language=Matlab,numbers=none,frame=tlbr,framesep=4pt,framerule=0pt]
[x, outs, opts] = solvePhaseRetrieval(A, At, b0, n, opts);
\end{lstlisting}
The user provides a measurement operator (as either a matrix or function handle) \texttt{A} and the data vector \texttt{b} as described in Equation \ref{S:1}. \texttt{At} is the Hermitian transpose/adjoint of \texttt{A}. If \texttt{A} is a function handle (as opposed to a matrix), then the user is required to provide \texttt{At} and also \texttt{n,} the length of the unknown signal vector. If \texttt{A} is a matrix, then the user can simply enter ``\texttt{[]}'' as an empty place-holder for these two arguments, and the implementation will infer them. \texttt{opts} is a struct containing options (see section \ref{S:22} and \ref{S:23} for details). 

The function returns 3 things:   An estimate of the unknown signal $\bm x$, a struct \texttt{outs} containing convergence and runtime information, and a struct \texttt{opts} containing a complete list of options that were used (including default values for options that were not set by the user).\\ 

\subsection{A worked example}
Here we demonstrate how to build and solve a problem using the Fienup method as an example.
First the user sets up the parameters and builds a test problem by invoking the helper function \path{buildTestProblem} to create a random Gaussian measurement matrix \texttt{A,} a random Gaussian signal \texttt{xt,} and a vector of measurements \texttt{b0.} In the test script \path{runSignalReconstructionDemo} we do this from scratch, but here we'll use the helper function to keep things short and simple.
\begin{lstlisting}[language=Matlab,numbers=none,frame=tlbr,framesep=4pt,framerule=0pt]
% Parameters
n = 100;          % Dimension of unknown vector
m = 5 * n;        % Number of measurements
isComplex = true; % If the signal and measurements are complex

%%  Build a random test problem
fprintf(`Building test problem...\n');
[A, xt, b0] = buildTestProblem(m, n, isComplex);

\end{lstlisting}


Then the user specifies the options in the struct \texttt{opts}.
\begin{lstlisting}[language=Matlab,numbers=none,frame=tlbr,framesep=4pt,framerule=0pt]
% Options
opts = struct;
opts.algorithm = `Fienup';
opts.initMethod = `truncatedSpectral';
opts.tol = 1e-10;
opts.maxIters = 500;
\end{lstlisting}
In this case, we've used the \texttt{opts.algorithm} and \texttt{opts.initMethod} options to select the solver (in this case, the Fienup method) and the initializer (the truncated spectral initialization method), respectively.  We set the desired accuracy at which the method terminates using \texttt{opts.tol}.  In this case, we've chosen an extremely high accuracy (1e-10, as opposed to something more standard like 1e-4). Because we've set such a strict tolerance, we also set \texttt{opts.maxIters} to control the maximum number of iterations allowed. A full list of options, algorithms and initializers are given in Section (\ref{S:22}). 

Now, we solve the phase retrieval problem using the specified options. 
\begin{lstlisting}[language=Matlab,numbers=none,frame=tlbr,framesep=4pt,framerule=0pt]
% Call the solver using the specified options
[x, outs, opts] = solvePhaseRetrieval(A, A', b0, n, opts);
\end{lstlisting}
Note, we handed \texttt{A'} and \texttt{n} to the solver.  This as actually not necessary. Because \texttt{A} is a dense matrix,  we could have simply entered ``\texttt{[]}'' for these arguments, and they would have been inferred.  If we were using a function handle instead of a matrix representation for \texttt{A}, these arguments would have been required.

Note that phase retrieval problems can only recover the signal up to an unknown phase ambiguity.  Next, we calculate the optimal rotation to align the recovered signal with the true signal, and calculate the relative reconstruction error.

\begin{lstlisting}[language=Matlab,numbers=none,frame=tlbr,framesep=4pt,framerule=0pt]
%% Determine the optimal phase rotation so that the recovered signal
%  matches the true signal as well as possible.  
alpha = (x'*xt)/(x'*x);
x = alpha * x;

%% Determine the relative reconstruction error. 
reconError = norm(xt-x)/norm(xt);
fprintf(`relative recon error = %d\n', reconError);
\end{lstlisting}
Finally, the user can plot the convergence results and the recovered signal \texttt{x} against the true signal \texttt{xt} using two helper functions.

\begin{lstlisting}[language=Matlab,numbers=none,frame=tlbr,framesep=4pt,framerule=0pt]
% Plot a graph of error versus the number of iterations.
plotErrorConvergence(outs, opts)

% Plot a graph of the recovered signal x against the true signal xt.
plotRecoveredVSOriginal(x,xt);
\end{lstlisting}

Further ready-made examples can be found in the \path{examples} folder within the PhasePack distribution.


\subsection{List of Algorithms, Options, and Initializers}
\label{S:22}

The following table lists the various algorithms available in Phasepack.   
\begin{longtable}  { | c | p{10cm} | }
 \hline
set  \texttt{opts.algorithm} = ...  & Description \\
 \hline
 \texttt{WirtFlow} & Wirtinger Flow Algorithm  \cite{candes2015phase}   \\
 \hline
  \texttt{TWF} & Truncated Wirtinger Flow Algorithm (with Poisson loss)  \cite{chen2015solving}   \\
\hline
 \texttt{RWF}  & Reweighted Wirtinger Flow Algorithm  \cite{Yuan:17}  \\
\hline
\texttt{AmplitudeFlow}   & Amplitude Flow Algorithm without truncation   \\
\hline
\texttt{TAF} & Truncated Amplitude Flow Algorithm    \cite{wang2016solving}\\
\hline
\texttt{RAF} & Re-Weighted Amplitude Flow Algorithm   \cite{2017arXiv170510407W} \\
\hline
\texttt{GerchbergSaxton}  & Gerchberg-Saxton Algorithm  \\
\hline
\texttt{Fienup}  & Fineup Algorithm  \cite{fienup1982phase} \\
\hline
\texttt{Kaczmarz}   & Kaczmarz Algorithm \cite{wei2015solving}  \\
\hline
\texttt{CoordinateDescent}  & Coordinate Descent Algorithm  \cite{2017arXiv170603474Z} \\
\hline
\texttt{PhaseMax}  & Phasemax Algorithm \cite{bahmani2016phase,goldstein2016phasemax}   \\
\hline
\texttt{PhaseLamp}  & Phaselamp Algorithm  \cite{dhifallah2017phase}  \\
\hline
\texttt{PhaseLift}  & Phaselift Algorithm \cite{candes2013phaselift,candes2015phase}  \\
\hline
\texttt{SketchyCGM} & SketchyCGM Algorithm  \cite{2017arXiv170206838Y}   \\
\hline
\end{longtable}

Each solver supports a range of options that customize the behavior of the method.   For example, Wirtinger flow and its variants can be made to run with non-linear conjugate gradient or L-BFGS acceleration by setting an appropriate value for \path{opts.searchMethod}  \cite{li2016gradient}.   The following commonly used options are supported by nearly all solvers. 
\begin{longtable}  { | c | p{10cm} | }
 \hline
 Option & Description \\
 \hline
 \texttt{opts.initMethod}  & string, default=`\texttt{orthogonal}'. The name of the initializer   \\
 \hline
 \texttt{opts.algorithm}  & string, default=`GerchbergSaxton'. The name of the algorithm.   \\
\hline
\texttt{opts.maxIters}  & integer, default=1000. Maximum number of iterations.   \\
\hline
\texttt{opts.maxTime}  & postive real number, default=120 (unit: seconds). Maximum time for the algorithm to run.  \\
\hline
\texttt{opts.xt}  & $n\times 1$ vector, default=[]. This is the true signal that generated the measurements. If provided, it will be used to compute reconstruction error, and will be used for determining stopping conditions.  In practice, \texttt{xt} is generally unknown, but \texttt{xt} can be supplied when solving synthetic problems for benchmarking and debugging.  \\
\hline
\texttt{opts.tol}  & real number, default=1.0e-6. The stopping tolerance.  Algorithms terminate when reconstruction error is less than \texttt{tol} if \texttt{xt} is provided.
                             Otherwise, terminate when residual is less than \texttt{tol}.  \\
\hline
\texttt{opts.verbose}  &  integer, default=0. \newline 0: print out nothing \newline 1: print out status information in the end. \newline 2: print out status information at every iteration.   \\
\hline
\texttt{searchMethod}  &  string, default=\path{steepestDescent}. Choose the type of gradient descent solver for Wirtinger flow type methods.  Valid options are \path{`steepestDescent'}, \path{`NCG'}, and \path{`LBFGS'}.  \\
\hline
\texttt{opts.recordMeasurementErrors}  & boolean, default=false. Whether to compute and record error (i.e. norm($|A\bm x|-\bm{b}$)/norm($\bm{b}$)) at each iteration.   \\
\hline
\texttt{opts.recordResiduals }  & (boolean, default=true) If  true, residual will be computed and recorded at each iteration. If false, 
residual won't be recorded. Residual also won't be computed if \texttt{xt} is provided. Note: The ``residual'' is an error metric used to measure convergence, and the definition varies across solvers.   \\
\hline
\texttt{recordReconErrors }  & boolean, default=false.  Whether to record reconstruction error. If it's true, \texttt{opts.xt} must be provided. If \texttt{xt} is provided, reconstruction error will be computed regardless of this flag and used for the stopping condition.  \\
\hline
\texttt{opts.recordTimes}  & boolean, default=true. Whether to record the runtime of each iteration. Total time will be measured regardless of this flag.  \\
\hline

\end{longtable}

The following table lists the various initializers available in Phasepack.
\newline

\begin{longtable}  { | c | p{11cm} | }
 \hline
 set  \texttt{opts.initMethod} =...  & Description \\
 \hline
 \texttt{Spectral} & Standard spectral Initializer   \cite{candes2015phase}  \\
 \hline
 \texttt{Truncated}& Truncated Spectral Initializer   \cite{chen2015solving}  \\
\hline
\texttt{Amplitude}  & Spectral initializer proposed for truncated amplitude flow \cite{wang2016solving}  \\
\hline
\texttt{Weighted} & Weighted spectral initializer \cite{2017arXiv170510407W}    \\
\hline
\texttt{Optimal} & Optimal (for random Gaussian measurements) spectral initializer \cite{mondelli2017fundamental}    \\
\hline
\texttt{Angle}  & Create a guess with specified angular distance to true signal   \\
\hline
\texttt{Orthogonal}  & Initial guess corresponding to maximum orthogonality with measurement vectors \cite{chen2015phase,wang2016solving}   \\
\hline
\end{longtable}

\subsection{Advanced Options}
\label{S:23}

In addition to general options, there are some algorithm-specific options that can be set.

\begin{center}
\begin{longtable}{|c|c|p{6cm}|}
\hline
\textsc{Algorithm} & \textsc{Options} & \textsc{Description} \\ \hline
\multirow{ 2}{*}{\texttt{Fienup}} & \texttt{opts.FienupTuning} & Tunning parameter for Fienup algorithm. 
                                          It influences the update of the fourier domain value 
                                           at each iteration.  \\
& \texttt{opts.maxInnerIters} &  The max number of iterations the inner-loop solver 
                                          will have.  \\ \hline

\multirow{ 2}{*}{\texttt{GerchbergSaxton}}
& \texttt{opts.maxInnerIters} &  The max number of iterations the inner-loop solver 
                                          will have.  \\ \hline

\texttt{Kaczmarz} & \texttt{opts.indexChoice} & Rule for choosing indices. Options are `cyclic' (default) and `random'  \\
\hline
\texttt{PhaseMax} & \texttt{opts.maxFastaIters} & Number of iterations that FASTA subroutine will have at each iteration  \\
\hline
\texttt{RAF} & \texttt{opts.reweightPeriod} & default = 20  \\
\hline
\texttt{SketchyCGM} & \texttt{opts.rank} & default = 1  \\
\hline
\multirow{ 2}{*}{\texttt{RWF}} & \texttt{opts.eta} & default = 0.9  \\
& \texttt{opts.reweightPeriod} &  default = 20 \\ \hline

\multirow{ 2}{*}{\texttt{TAF}} & \texttt{opts.gamma} & default = 0.7  \\
& \texttt{opts.truncationPeriod} &  default = 20 \\ \hline

\hline

\end{longtable}
\end{center}


\section{Benchmarking \& Comparing Solvers}
\label{benchmark}

Algorithms can be compared against each other to produce plots including those shown in Figure \ref{fig:random}. This is done by invoking the general benchmark interface:
\begin{lstlisting}[language=Matlab,numbers=none,frame=tlbr,framesep=4pt,framerule=0pt]
benchmarkSynthetic(xitem, xvals, yitem, algorithms, dataSet, params);
\end{lstlisting}
Note that this benchmarking routine supports both real and synthetic measurement matrices.  This routine produces measurements by applying a measurement matrix (either Gaussian, Fourier, or a real transmission matrix measured using an optical apparatus) to a known ground-truth signal, and then contaminating the results with noise.  The signal is then reconstructed and the accuracy is measured.  Reconstructions are done with different algorithms, and under varying conditions, and performance curves are produced.

Each of the arguments to the benchmark routine are described below.
\begin{itemize}[leftmargin=3em]
  \item \texttt{xitem}: String that indicates the independent variable that is varied over the benchmark. The possible choices include \texttt{`m/n'}, \texttt{`snr'}, \texttt{`masks'} (the number of Fourier masks), \texttt{`iterations'}, \texttt{`time'} (total allowed runtime), and \texttt{`angle'} (angular distance between true solution and initializer).   An example of how to set this argument is shown below.  See Section \ref{benchTable} for a complete list of options.
\begin{lstlisting}[language=Matlab,numbers=none,frame=tlbr,framesep=4pt,framerule=0pt]
% Choose values to be shown on the x-axis
xitem = `m/n'; 			  % Vary the oversampling ratio on the x-axis 
\end{lstlisting}

In this case, \texttt{m/n} refers to the ratio of the number of measurements to the signal length.

\item \texttt{xvals}: A list containing the values that \texttt{xitem} will assume.
 \begin{lstlisting}[language=Matlab,numbers=none,frame=tlbr,framesep=4pt,framerule=0pt]
xvals = [2 4 6 8]; 	 % Run trials for these values of m/n
\end{lstlisting}

\item \texttt{yitem}: String indicating the metric used to measure the performance of the selected algorithms. The possible choices are \texttt{`reconError'}, \texttt{`measurementError'}, and \texttt{`correlation'}.   See Section \ref{benchTable}.
\begin{lstlisting}[language=Matlab,numbers=none,frame=tlbr,framesep=4pt,framerule=0pt]
% Choose the metric to be evaluated for each method, and plotted on the y-axis
yitem = `reconError';   
\end{lstlisting}

\texttt{reconError} corresponds to the reconstruction error. It is the relative difference (as measured by the 2-norm) between the true signal and the reconstructed signal.

  \item \texttt{algorithms}: A cell array of structs that enumerates the algorithms to be benchmarked. For example, to run a comparison between the Fienup and the truncated amplitude flow algorithms, one would build a struct for each method like this:
\begin{lstlisting}[language=Matlab,numbers=none,frame=tlbr,framesep=4pt,framerule=0pt]
fienup = struct(`algorithm',`fienup');    
taf = struct(`algorithm',`taf');
algorithms = {fienup, taf};
\end{lstlisting}

\item \texttt{dataSet}: An option for choosing whether to test on synthetic or real world measurement matrices. Options include `1DGaussian', `2DImage', and `transmissionMatrix'. See Section \ref{tab:data}.
\begin{lstlisting}[language=Matlab,numbers=none,frame=tlbr,framesep=4pt,framerule=0pt]
% Test using random Gaussian measurements
dataSet = `1DGaussian'; 
\end{lstlisting}

\texttt{1DGaussian} refers to a synthetic Gaussian random matrix.  PhasePack also supports testing on real-world measurement matrices by using the `\path{transmissionMatrix}' option. 

\item \texttt{params}: General parameters that can effect output and experiments when required. For example, if the user wants to see convergence information print after each iteration of the algorithms, they would set a flag as follows. 

\begin{lstlisting}[language=Matlab,numbers=none,frame=tlbr,framesep=4pt,framerule=0pt]
params.verbose = true;
\end{lstlisting}

A full set of parameters is shown in Section (\ref{tab:params}).
\end{itemize}

\subsection{A more complex example}

Here we take \texttt{runBenchmark1DGaussianMN} as an example to demonstrate the usage of the benchmark interface. It compares several algorithms with each other at different $m/n$ ratios (where $m$ is the number of measurements and $n$ is the signal length).\\

First, the user sets up the metrics over which the algorithms will be compared with one another.  In this case, we look at how the number of measurements effects performance, and we choose which values the $m/n$ ratio will assume.
\begin{lstlisting}[language=Matlab,numbers=none,frame=tlbr,framesep=4pt,framerule=0pt]
% Choose x label, values shown on the x axis and y label
xitem = `m/n';
xvalues = [1 2 2.25 2.5 2.75 3 3.25 3.5 3.75 4 5 6]; 
yitem = `reconerror';
\end{lstlisting}

Then the user chooses the dataset for benchmarking. In this case, we choose 1-D Gaussian data.
\begin{lstlisting}[language=Matlab,numbers=none,frame=tlbr,framesep=4pt,framerule=0pt]
% Choose Dataset and set up parameters
dataSet = `1DGaussian';
\end{lstlisting}

Next, the user creates the parameters struct. The comments highlighted in green indicate the description for each parameter.
\begin{lstlisting}[language=Matlab,numbers=none,frame=tlbr,framesep=4pt,framerule=0pt]
params.verbose = false;		% keep text output short and readable
params.numTrials = 5;     % run several random trials for each scenario, and report average results
params.n = 20;            % number of unknown elements
params.isComplex = true;  % use complex matrices? or just stick to real?
params.policy = `median'; % use the median performance over trails to assess algorithms


\end{lstlisting}

Finally, the user creates a set of structs for all the algorithms that are to be compared. 

\begin{lstlisting}[language=Matlab,numbers=none,frame=tlbr,framesep=4pt,framerule=0pt]
wf = struct(`initMethod',`spectral',`algorithm',`wirtflow');
twf = struct(`algorithm',`twf'); 
rwf = struct(`algorithm',`rwf');
ampflow = struct(`algorithm',`amplitudeflow');
taf = struct(`initMethod',`orthogonal',`algorithm',`taf');
raf = struct(`initMethod',`weighted',`algorithm',`raf');
fienup = struct(`algorithm',`fienup');
gs = struct(`algorithm',`gerchbergsaxton');
cd = struct(`algorithm',`coordinatedescent',`maxIters',3000);
kac = struct(`algorithm',`kaczmarz',`maxIters',1000);
pmax = struct( `algorithm',`phasemax', `maxIters',1000);
plamp = struct(`algorithm', `phaselamp');
scgm = struct(`algorithm',(`sketchycgm');     
plift = struct(`algorithm',(`phaselift',(`maxIters',1000);     

% Grab your pick of algorithms.
algorithms = {gs,wf,plift,pmax,plamp};
\end{lstlisting}

Finally, we call the benchmark routine.
\begin{lstlisting}[language=Matlab,numbers=none,frame=tlbr,framesep=4pt,framerule=0pt]
benchmarkSynthetic(xitem, xvalues, yitem, algorithms, dataSet, params);
\end{lstlisting}

Note that unlike the single solver example, one does not need to either manually compute the reconstruction errors or plot the convergence results using helper functions. The function \texttt{benchmarkSynthetic} automatically computes the errors, plots all the curves in a single plot, and displays it at the end of the run.\\

A number of complete ready-made examples can be found in the \texttt{benchmarks} folder.

\subsection{List of Datasets, Parameters, and Options for Benchmarking}
\label{benchTable}
This section lists the possible choices for \texttt{datasets}, \texttt{xitem}, \texttt{yitem}, and the general parameters \texttt{params}.\\

The following is a list of possible datasets, their short description, and the \texttt{xitem} choices that each supports.  Note:  before using the empirical datasets (TM16, TM40, TM64), these datasets must be downloaded and added to the \path{data} folder.  See Section \ref{sec:tm} for instructions.
\begin{center}
\label{tab:data}
\begin{longtable}  {| c | p{8cm} | p{4cm} |} 
 \hline
\texttt{dataSet=...} & description & supported \texttt{xitem}'s\\    
 \hline
 \texttt{`1DGaussian'}  & Both measurement matrix $A$ and the true signal $\bm{x}$ are generated from 1D random Gaussian distribution. & \texttt{m/n}, \texttt{snr}, \texttt{iterations}, \texttt{time}\\
 \hline
 \texttt{`2DImage'}  &  Synthetic data are generated from a specified image using a specified number of octanary masks. The image is specified using \texttt{params.imagePath}. & \texttt{masks}, \texttt{snr}, \texttt{iterations}, \texttt{time}\\
\hline
\texttt{`TM16',`TM40',`TM64'}  & TM 16 x 16 amplitude dataset, TM 40 x 40 phase dataset and TM 64 x 64 amplitude dataset. See Section \ref{sec:tm}  for details.  &  \texttt{m/n}, \texttt{snr}, \texttt{iterations}, \texttt{time}\\
\hline
\end{longtable}
\end{center}

The following is a list of possible options for the x-axis.\\
\begin{center}
\label{tab:xitem}
\begin{longtable}  {| c | p{14cm} |} 
 \hline
\texttt{xlabel=...} & DESCRIPTION\\    
 \hline
 \texttt{`m/n'}  & The ratio between the length of  the data vector $\bm b$ and the length of the unknwon signal $\bm x$.\\
\hline
\texttt{`masks'}  &  The number of octanary Fourier masks used for acquiring data.  This option is only used for the \texttt{`2DImage'} dataset.  For other datasets, use the `m/n' option to control the number of measurements. \\
\hline
\texttt{`snr'}  &  The signal-to-noise ratio.\\
\hline
\texttt{`iterations'}  &  The number of iterations allowed before termination.\\
\hline
\texttt{`time'}  & The max running time for a single trial (unit: seconds).\\
\hline
\end{longtable}
\end{center}

The following is a list of possible options for the y-axis.\\
\begin{center}
\label{tab:ylab}
\begin{longtable}  { | c | p{12cm} | }
 \hline
\texttt{yitem=...} & DESCRIPTION \\    
 \hline
 \texttt{`reconError'}  & The relative 2-norm difference between the true and reconstructed image.\\
\hline
\texttt{`measurementError'}  & The relative 2-norm difference between the original measurements and measurements extracted from the recovered signal. \\
\hline
\texttt{`correlation'}  & Correlation between true and recovered signal.\\
\hline
\end{longtable}
\end{center}

\newpage
The following is a list of possible options for the general parameters.
\begin{center}
\label{tab:params}
\begin{longtable}  { | c | p{12cm} | }
 \hline
parameter field & description \\    
 \hline
 \texttt{params.verbose}  & boolean, default=false. If true, the result of each trial will be reported.   \\
 \hline
\texttt{params.numTrials}  & integer, default=1. The number of random trials to run at each \texttt{xvalue}.  Reported results are averaged over this many random trials.  \\
\hline
\texttt{params.policy}  & string, default='median'.  How to compute the final value used for plotting.  Option include:
\begin{itemize}
\item \texttt{`average'}: Take the average of all trials.
\item \texttt{`best'}: Take the best result of all trials. 
\item \texttt{`successrate'}: Report the number of times that exact signal reconstruction took place.  Reconstruction is considered exact when \texttt{ylabel} is less than \texttt{params.successConstant}.
\end{itemize}
\\
\hline
\texttt{params.maxTime}  &  real number, default=120 (unit: seconds). Maximum time for each trial to run.  \\
\hline
\texttt{params.recordSignals}  & boolean, default=false. Whether to record the recovered signal after each trial.\\
\hline
\end{longtable}
\end{center}

\section{Using empirical datasets}
Phasepack uses the empirical datasets described in \cite{metzler2017coherent}.  These datasets consists of measurement matrices for the task of imaging through a diffusive medium.  Measurement matrices are provided for acquiring images of resolutions $16\times 16$, $40\times 40$, and $64\times 64.$ These matrices are stored in dense matrix form, and so computing at the higher resolutions is very computation and memory intensive.
 For each image resolution, there are 5 sets of empirical phaseless measurements, each corresponding to a different image.  These 5 images can be reconstructed by applying phase retrieval to these data.  
 
 For a complete description of these datasets and how they were acquired, see  \cite{metzler2017coherent}.

\subsection{Getting empirical datasets}\label{sec:tm}

Before using the empirical transmission matrix datasets, the data needs to be downloaded and installed into the \path{data} folder.
To do this, visit \path{https://rice.app.box.com/v/TransmissionMatrices/}, and scroll down to the hyperlink for downloading the dataset.  Then, click the black ``Download'' button in the upper right corner. The download is a zip file named `TransmissionMatrices.zip'.  The folder contains over 1Gb of data, and may take some time to download.

 Unzip the folder, and place the contents in the location \path{data/TransmissionMatrices} within the main PhasePack folder.  If successful, \path{data/TransmissionMatrices} should contain the sub-folders \texttt{Coherent\_Data} and \texttt{Functions}, and the file \texttt{ReadMe.txt}.

For details of this phase retrieval problem, and how the transmission matrix data is constructed, see \cite{metzler2017coherent}.

\subsection{Benchmarking with empirical data}
\sloppy
In Section \ref{benchmark}, we describe how to benchmark methods using synthetic signals.  The methods described in that Section can be use with the \texttt{transmissionMatrix} dataset, which uses empirical measurement matrices with synthetic signals (so that a ground truth baseline is known).  

It is also possible to compare methods using both empirical measurement matrices and empirical data using the \texttt{benchmarkTransmissionMatrix} interface. This method will load a transmission matrix and phaseless measurements from a data file, reconstruct an image from the measurements, and report error for different algorithms.  Sample results are shown in Figure \ref{fig:emp}.

An example of how to use this benchmark method is shown in the file \texttt{runBenchmarkTM16MN.m}.  First, we choose the dataset to use.  The transmission matrix dataset comes with measurement matrices at 3 different image resolutions ($16\times 16$, $40\times 40$, $64\times 64$), and there are 5 different sets of measurements (corresponding to different image targets) at each resolution.  
\begin{lstlisting}[language=Matlab,numbers=none,frame=tlbr,framesep=4pt,framerule=0pt]
%% 1.Set up parameters
imageSize = 16;         % Side-length of image to reconstruct. Valid choices: 16,40,64
datasetSelection = 4;   % Which dataset to select measurements from.  Valid choices: 1-5
residualConstant = 0.4; % Cutoff to load only high-quality rows of the transmission matrix.  
\end{lstlisting}
The rows of the measurement matrix were calculated using a measurement process (also a phase retrieval problem, see \cite{metzler2017coherent}), and some are more accurate than others.  Each measurement matrix comes with a per-row ``residual'' to measure the accuracy of that row.  The \texttt{`residualConstant'} is a cutoff to throw out inaccurate rows; only rows with a residual below this cutoff will be loaded. 

Finally, we select the algorithm to run this benchmark on, and call the benchmark routine.
\begin{lstlisting}[language=Matlab,numbers=none,frame=tlbr,framesep=4pt,framerule=0pt]
% Create a list of algorithms structs
wf = struct(`initMethod',(`spectral',(`algorithm',(`wirtflow');
twf = struct(`algorithm',`twf'); 
rwf = struct(`algorithm',`rwf');
ampflow = struct(`algorithm',`amplitudeflow');
taf = struct(`initMethod',`orthogonal',`algorithm',`taf');
raf = struct(`initMethod',`weighted',`algorithm',`raf');
fienup = struct(`algorithm',`fienup');
gs = struct(`algorithm',`gerchbergsaxton');
cd = struct(`algorithm',`coordinatedescent',`maxIters',30000);
kac = struct(`algorithm',`kaczmarz',`maxIters',1000);
pmax = struct( `algorithm',`phasemax', `maxIters',1000);
plamp = struct(`algorithm', `phaselamp');
scgm = struct(`algorithm',`sketchycgm');     
plift = struct(`algorithm',`phaselift',`maxIters',1000);                                             

% Grab your pick of algorithms.
algorithms = {raf,gs};

%% 2. Run benchmark
benchmarkTransmissionMatrix(imageSize, datasetSelection, residualConstant, algorithms)
\end{lstlisting}
As the benchmark runs, images will be reconstructed.  These images are saved in the \path{benchmarkResults} directory.

\section{Adding new solvers}

To add a new solver to PhasePack, first create a new function in the \path{solvers} folder with a signature like the following example.
\begin{lstlisting}[language=Matlab,numbers=none,frame=tlbr,framesep=4pt,framerule=0pt,frame=tlbr,framesep=4pt,framerule=0pt]
function [sol, outs] = myNewSolver(A, At, b0, x0, opts)
\end{lstlisting}
The function should treat the operators \texttt{A} and \texttt{At} as function handles (i.e., they are invoked by calling \texttt{A(x)} and \texttt{At(x)}).   The function should return the solution to the phase retrieval problem described by \texttt{A} and \texttt{b0}, in addition to the struct \texttt{outs} containing any relevant convergence information. 

Choose a string that will be used to choose the new algorithm when calling \texttt{solvePhaseRetrieval}.  In this example, we'll choose the string ``mysolver.''
 This argument is not case sensitive (PhasePack automatically converts the string to lower case), so the user can choose the solver by using the string ``mySolver.''  However, within the internal PhasePack code, we always use lower case characters.

Next, edit the file \path{utils/manageOptions}.  Inside that file, the method \texttt{getDefaultOpts} returns a struct with the default options for a method.  Add a \texttt{case} to the \texttt{switch} statement that sets any default values for options that you want to have available.  
\begin{lstlisting}[language=Matlab,numbers=none,frame=tlbr,framesep=4pt,framerule=0pt,frame=tlbr,framesep=4pt,framerule=0pt]

function opts = getDefaultOpts(algorithm)
	...											% Code specifying various default options 
	switch lower(algorithm)
	...							% Code specifying various algorithm-specific default options 
		case 'mysolver'.      		% Put options for your new algorithm here!
			opts.maxIters = 1000;
			opts.myAlgorithmSpecificParameter = .1;
	...
\end{lstlisting}

Finally, inside the file \path{solvers/solvePhaseRetrieval} find the function \texttt{solveX}.  This function contains a switch statement that invokes the solver corresponding to an algorithm name.  Add an entry to the switch statement that invokes your new algorithm when the appropriate string is given as the algorithm name.  Remember, the name used in the switch statement must be all lower case.

\begin{lstlisting}[language=Matlab,numbers=none,frame=tlbr,framesep=4pt,framerule=0pt,frame=tlbr,framesep=4pt,framerule=0pt]
function [sol, outs] = solveX(A, At, b0, x0, opts)
	switch lower(opts.algorithm)
	...													% Code for calling other solvers
		case 'mysolver'						% Call your new solver here!
			[sol, outs] = myNewSolver(A, At, b0, x0, opts);
	...
\end{lstlisting}

And that's it.  You can now use your algorithm like this.

\begin{lstlisting}[language=Matlab,numbers=none,frame=tlbr,framesep=4pt,framerule=0pt]
...															% Setup your problem
opts.algorithm = `mysolver'			% Choose your new solver
[x, outs, opts] = solvePhaseRetrieval(A, At, b0, n, opts);
\end{lstlisting}

\newpage
\section{References}
\label{S:5}
\bibliographystyle{plain}
\bibliography{references}





\end{document}